\documentstyle[aps, multicol]{revtex}
\author{Ernesto Fuenmayor$^a$\footnote{efuenma@fisica.ciens.ucv.ve},
Lorenzo Leal$^a$\footnote{lleal@fisica.ciens.ucv.ve} and Ryan
Revoredo$^{a,b}$\footnote{revoredoryan@yahoo.com}}
\address{$^a$Grupo de Campos y Part\'{\i}culas, Departamento de
F\'{\i}sica, Facultad de Ciencias, Universidad Central de Venezuela, AP
47270, Caracas 1041-A, Venezuela\\
$^b$ Departamento de Matem\'atica, Universidad Metropolitana, Caracas,
Venezuela}
\title{Loop Representation of charged particles interacting with Maxwell
and Chern-Simons fields}

\begin{document}

\maketitle
\begin{abstract}
 The loop representation formulation of non-relativistic particles
coupled with abelian gauge fields is studied. Both Maxwell and
Chern-Simons interactions are separately considered. It is found that
the loop-space formulations of these models share significant
similarities, although in the Chern-Simons case there exists an unitary
transformation that allows to remove the degrees of freedom associated
with the paths. The existence of this transformation, which allows to
make contact with the anyonic interpretation of the model, is subjected
to the fact that the charge of the particles be quantized. On the other
hand,  in the Maxwell case, we find that charge quantization is
necessary in order to the geometric representation be consistent.
\end{abstract}
\begin{multicols}2
\section{Introduction}
The loop representation (L.R.) constitutes an useful tool in present day
investigations in gauge theories \cite{O-1,O-2}. There are several
approaches to the L.R. \cite{1a,1b,1c,1d}, all of them sharing the
recognition of string-like structures as the basic objects needed to
build a geometric representation for gauge field quantization.

In this paper we study the L.R. formulation of point particles
interacting with abelian gauge fields. The coupling of point particles
to fields presents certain subtleties that make the canonical
quantization far from being straightforward. In turn, the corresponding
L.R. shows its own particularities, which had not yet been reported.
This study is carried out first for non-relativistic dynamical point
particles in electromagnetic interation. We shall not worry about the
lack of Lorentz covariance, neither we shall discuss regularization
issues. As we shall see, for the L.R. formulation of this model to be
consistent, charge must be quantized. This result should be compared
with a similar one obtained several years ago for the Maxwell theory,
within the Spin Networks version of the L.R. \cite{3,3a}.

As a second model we consider the topological interaction between
non-relativistic dynamical charged particles caused by a Chern-Simons
term \cite{4,5}. Althought both theories share the same geometrical
framework when quantized in the L.R., in the Chern-Simons case the loop
dependence may be eliminated by means of an unitary transformation,
which yields a quantum mechanics of many particles subjected to a long
range interaction. As we shall discuss, this unitary transformation
holds provided charge is quantized.

This paper is organized as follows. In section II we study the L.R.
formulation of non-relativistic point particles in electromagnetic
interaction. Section III is devoted to consider the L.R. quantization
of  point particles with Chern-Simons interaction. Some final remarks
are left for the last section.

\section{Electromagnetic interaction of non-relativistic point
particles}
The action for $N$ electromagnetically  interacting non-relativistic
charged particles may be writen as

\begin{eqnarray}
S&=&\int
dt\sum_{p=1}^{N}\left(\frac12\,m_{(p)}|\dot{\vec{r}}_{(p)}|^2-e\,q_{(p)}\dot{r}^{i}A_{i}(\vec{r}_{(p)},t)\right.\nonumber\\
        &&\left.-e\,q_{(p)}A_{o}(\vec{r}_{(p)},t)\right)-\frac{1}{4}\int
d^{4}x \,F^{\mu\nu}(\vec{x},t)F_{\mu\nu}(\vec{x},t)\;,
\label{ec1}
\end{eqnarray}
where $F_{\mu\nu}=\partial_{\mu}A_{\nu}-\partial_{\nu}A_{\mu}$ and
$\vec{r}_{(p)}$, $q_{(p)}$ denote the position and charge of the $p-th$
particle respectively.

After Dirac quantization in the $A_{o}=0$ gauge, one obtains the first
class Hamiltonian

\begin{eqnarray}
H&=&\sum_{p=1}^{N}\frac{1}{2m_{(p)}}\left(p_{(p)i}+eq_{(p)}A_{i}(\vec{r}_{(p)},t)\right)^{2}\;\nonumber\\
&&+ \int d^{3}x \;\frac{1}{2}\left(|\vec{E}|^{2}+|\vec{B}|^{2}\right)\;,
\label{ec2}
\end{eqnarray}
together with the Gauss (first class) constraint:

\begin{equation}
\varphi\equiv\partial_{i}E^{i}-\sum_{p}e\,q_{(p)}\delta^{3}(\vec{r}_{(p)}-\vec{x})\approx0\;.
\label{ec3}
\end{equation}
In these equations $e$ is the electromagnetic coupling constant (which
in $3+1$ space is dimensionless), while $E^{i}\equiv F^{io}$ and
$B^{i}=-\frac{1}{2}{\epsilon}^{ijk}F_{jk}$ denote the electric and
magnetic fields. The operators $\vec{r}_{(p)}$ and $\vec{p}_{(p)}$ are
canonical conjugates, likewise $\vec{A}$ and $\vec{E}$:

\begin{eqnarray}
\left[r^{i}_{(p)},p_{(q)j}\right]&=&i\delta^{i}_{j}\delta_{pq}\;,
\label{dificil}
\end{eqnarray}

\begin{eqnarray}
\left[A_{i}(\vec{x}),E^{j}(\vec{y})\right]&=&i\delta^{j}_{i}\delta^{3}(\vec{x}-\vec{y})\;.
\label{ec4}
\end{eqnarray}
The expression $A_{i}(\vec{r}_{(p)},t)$ is a shorthand for

\begin{equation}
A_{i}(\vec{r}_{(p)},t)\equiv \int
d^{3}\vec{x}\;\delta^{3}(\vec{x}-\vec{r}_{(p)})A_{i}(\vec{x},t)\;,
\label{ec5}
\end{equation}
where $\delta^{3}(\vec{x}-\vec{r}_{(p)})$ is an operator-valued
distribution acting on the Hilbert space of the $p$-th particle:

\begin{equation}
\delta^{3}(\vec{x}-\vec{r}_{(p)Operator})\,|\vec{r}_{(p)}\rangle =
\delta^{3}(\vec{x}-\vec{r}_{(p)})\,|\vec{r}_{(p)}\rangle\;.
\label{ec6}
\end{equation}
The full Hilbert space of the theory may be spanned by the basis $\prod
|\vec{r}_{(p)}\rangle \otimes |\vec{A}\rangle$, constructed by taking
the tensorial product of the ``position'' eigenstates
$|\vec{r}_{(p)}\rangle$ and $|\vec{A}\rangle$ associated to the
particles and the field respectivelly. The Hilbert space must be
restricted to the physical space, in Dirac sense, defined by $\varphi\,
|\psi_{Physical}\rangle =0$. Also, we must identify which operators are
first class, i.e., gauge invariant [remember that the Gauss constraint
(\ref{ec3}) generates spatial gauge transformations, both on particle
and field operators]. It is inmediate to check that the electric and
magnetic fields $\vec{E}$, $\vec{B}$, together with the particle
position operator $\vec{r}_{(p)}$ and the gauge covariant momentum
$\vec{p}_{(p)}+e\,q_{(p)}\vec{A}(\vec{r}_{(p)},t)$ commute with the
Gauss constraint, unlike the gauge dependent operators $\vec{p}_{(p)}$
and $\vec{A}$. It is worth mentioning that every physical observable may
be constructed in terms of the first class operators mentioned above
[see, for instance, expression (\ref{ec2}) for the energy of the
field-particles system].

Next, let us consider the L.R. appropriate to the theory we are dealing
with. A brief review of how it works in the sourceless case will help.
In the pure Maxwell theory \cite{1a,1b,1c,1d}, the gauge invariante
operators $\vec{E}$ and $\vec{B}$ may be realized onto loop dependent
wave funtionals $\Psi (C)$ as:

\begin{equation}
E^{i}\,\Psi (C)=e\,T^{i}(\vec{x},C)\,\Psi (C)\;,
\label{ec7}
\end{equation}

\begin{equation}
F_{ij}\,\Psi (C)=i/e\,\Delta_{ij}(\vec{x})\,\Psi (C)\;,
\label{ec8}
\end{equation}
where the form factor

\begin{equation}
T^{i}(\vec{x},C)\equiv \oint dy^{i} \, \delta^{3}(\vec{x}-\vec{y})\;,
\label{ec9}
\end{equation}
is a distributional vector density that encodes the information of the
shape of the spatial loop $C$. The loop derivative of Gambini-Tr\'{\i}as
$\Delta_{ij}(\vec{x})$ \cite{1a,1b,1c,1d} is defined as:

\begin{equation}
\Psi (\sigma \cdot C)= \left(1+ \sigma^{ij}\Delta_{ij}(\vec{x})\right)\,
\Psi (C)\;,
\label{ec10}
\end{equation}
with $\sigma^{ij}$ being the area of an infinitesimal plaquette attached
at the spatial point $\vec{x}$. Thus $\Delta_{ij}(\vec{x})$ measures how
the loop dependent funtion $\Psi (C)$ changes under a small deformation
of its argument $C$. In the loop representation, the source-free Gauss
law  constraint $(\partial_{i}E^{i}=0)$ is automatically satisfied,
since $T^{i}(\vec{x},C)$ has vanishing divergence. One can thus
interpret $C$ as a closed Faraday's line of electric flux.

In the case of particles interacting with fields, one needs to
enlarge the space of states. To simplify the discussion, let us
begin by considering the one particle case. The interpretation of
loops as Faraday's lines of electric flux, leads in a natural way
to try the following picture: consider an open path
$\gamma_{\vec{r}}$ starting at the particle's position $\vec{r}$
and ending at the spatial infinity [to take into account the
source-free sector, this open path might be accompanied by closed
contours too]. Then, consider path-dependent wave functionals
$\Psi (\gamma_{\vec{r}})$, and define the action of the electric
field operator as in equation (\ref{ec7})

\begin{equation}
E^{i}(\vec{x})\,\Psi
(\gamma_{\vec{r}})=e\,T^{i}(\vec{x},\gamma_{\vec{r}})\,\Psi
(\gamma_{\vec{r}})\;.
\label{ec12}
\end{equation}
Then, the Gauss constraint (\ref{ec3}) states that:

\begin{eqnarray}
\big(&e&\partial_{i}T^{i}(\vec{x},\gamma_{\vec{r}})-e\,q\,\delta^{3}(\vec{r}-\vec{x})\;\big)\,\Psi(\gamma_{\vec{r}})\nonumber\\
&=&e\left(\delta^{3}(\vec{r}-\vec{x})-q\,\delta^{3}(\vec{r}-\vec{x})\right)\,\Psi(\gamma_{\vec{r}})\nonumber\\
&=&0,
\label{ec13}
\end{eqnarray}
where we have dropped the $\delta^{3}(\infty)$ contribution
arising from the end of the path. Equation (\ref{ec13}) implies
that $q=1$. This result provides the key to complete the picture
of the kinds of paths allowed. Had we taken an incoming path
instead of the outgoing one, the Gauss law had been satisfied
only for $q=-1$. On the other hand, if we take a ``multiple''
open path, i.e., $n$ strands outgoing (incoming) from (towards)
$\vec{r}$, the allowed value for $q$ would be $n$ $(-n)$.
Finally, it is easy to see that for $N$ charges, one must take
$N$ ``bundles'' of open paths, one for each charge $q_{(p)}$,
having as many strands as the value of the charge, and oriented
according to its sign. Hence, within this formalism there is no
room for fractionary charges: a Faraday line carries one unit of
electric flux $e$, which must be emitted from or absorbed by an
integral charge $q_{(p)}$. Then one has:

\begin{equation}
\sum_{p=1}^{N}q_{(p)}\delta^{3}(\vec{x}-\vec{r}_{(p)})=
\sum_{s}\left(\delta^{3}(\vec{x}-\vec{a}_{s})-\delta^{3}(\vec{x}-\vec{b}_{s})\right)\;,
\label{ec14}
\end{equation}
with $\vec{a}_{s}$ and $\vec{b}_{s}$ labeling the starting and ending
points of the $s$-th ``strand'', and the Gauss constraint (\ref{ec3})
becomes an identity on the physical states.

It remains to study whether or not the algebra of observables admits a
realization in terms of operators acting on these path-dependent
(Faraday's lines dependent) functionals $\Psi (\gamma_{\vec{r}})$.
Besides the electric and magnetic fields, wich are realized as in
equations (\ref{ec7}) and (\ref{ec8}) [remember that the paths may also
be comprised by closed loops, hence the loop derivative makes sense in
this context too], we prescribe:

\begin{eqnarray}
&&p_{(p)i}+e\,q_{(p)}\,A_{i}(\vec{r}_{(p)},t)\rightarrow\nonumber\\
&&-iD_{i}(\vec{r}_{(p)})\equiv -i\left(\frac{\partial}{\partial
r_{(p)}^{i}}-q_{(p)}\,\delta_{i}(\vec{r}_{(p)})\right)\;,
\label{ec15}
\end{eqnarray}
where $\delta_{i}(\vec{x})$ is the ``path derivative'', that acts onto
path-dependent functions $\Psi(\gamma_{\vec{r}_{(p)}})$ by measuring
their change when an infinitesimal open path starting at $\vec{x}$ and
ending at $\vec{x}+\vec{h}$ $(\vec{h}\rightarrow 0)$ is appended to the
list of paths comprised in $\gamma_{\vec{r}_{(p)}}$ \cite{8}:

\begin{equation}
\Psi(h\cdot
\gamma_{\vec{r}_{(p)}})=\left(1+h^{i}\,\delta_{i}(\vec{r}_{(p)})\right)\,\Psi(\gamma_{\vec{r}_{(p)}})\;.
\label{ec16}
\end{equation}

The $\delta_{i}(\vec{x})$ derivative is related with the loop derivative
(\ref{ec10}) through:

\begin{equation}
\Delta_{ij}(\vec{x})= \frac{\partial}{\partial
x^{i}}\delta_{j}(\vec{x})-\frac{\partial}{\partial
x^{j}}\delta_{i}(\vec{x})\;.
\label{ec17}
\end{equation}

The gauge invariant combination $D_{i}(\vec{r}_{(p)})$ coincides with
the derivative introduced by Mandelstam several years ago \cite{9}. It
comprises the ordinary derivative, representing the momentum operator of
the particle, plus $q_{(p)}$ times the ``path derivative''
$\delta_{i}(\vec{r}_{(p)})$. The ``Mandelstam operator''
$D_{i}(\vec{r}_{(p)})$ has a nice geometric interpretation within the
present formulation, as we shall see. In this representation, both
particles and fields are described by geometric means: particles are
labelled by points $\vec{r}_{(p)}$ (as usual), and fields by open paths.
Gauge invariance restricts paths to be closed, or to start (or end) at
the points where particles ``live''. Gauge invariant operators, on the
other hand, respect the geometrical properties dictated by gauge
invariance: the ``position'' operators $\vec{r}_{(p)}$ and $\vec{E}$,
are diagonal in this representation, and act by displaying the
localization and shape of the geometric configurations. In turn, the
magnetic field operator computes the change in the wave functional when
a small ``plaquette'' is added, while the covariant momentum
$-iD_{i}(\vec{r}_{(p)})$ measures the change when both the particle and
its attached ``bundle'' of paths are infinitesimally displaced. In both
cases, the involved derivative operation fulfills the geometrical
requirements imposed by gauge invariance. At this point, it should be
observed that a more appropriate notation for the path dependent
functionals would be $\Psi(\gamma_{\vec{r}_{(p)}}, \vec{r}_{(p)})$,
since it displays both the path and point-dependence, which are affected
by the path and ordinary derivatives respectivelly.

Finally, it can be shown that the path-space operators obey the algebra
arising from the canonical commutators, i.e., they constitute a
representation of the quantum theory under study. For instance, one has

\begin{eqnarray}
&&\left[-iD_{i}(\vec{r}_{(p)}),-iD_{j}(\vec{r}_{(p)})\right]\Psi(\gamma_{\vec{r}_{(p)}},
\vec{r}_{(p)})\nonumber\\
&&=q_{(p)}\,\Delta_{ij}(\vec{r}_{(p)})\Psi(\gamma_{\vec{r}_{(p)}},
\vec{r}_{(p)})\;,
\label{ec18}
\end{eqnarray}
which corresponds to the relation

\begin{eqnarray}
&&\left[p_{(p)i}+e\,q_{(p)}\,A_{i}(\vec{r}_{(p)}),
\,p_{(p)j}+e\,q_{(p)}\,A_{j}(\vec{r}_{(p)})\right]\nonumber\\
&&=-ie\,q_{(p)}\,F_{ij}(\vec{r}_{(p)}) \;.
\label{ec19}
\end{eqnarray}

Summarizing, we saw that the L.R. of the Maxwell theory coupled
with point charged particles is a ``Faraday's lines
representation'' that may be set up only if electric charges are
quantized, the fundamental unit of charge being the
electromagnetic coupling constant $e$, which in this framework is
the unit of electric flux carried by each Faraday's line.

\section{Non-relativistic point particles interacting through
Chern-Simons field}
We now turn our attention to the model described by the action,

\begin{eqnarray}
S&=&\int dt \sum_{p=1}^{N}
\Big[\frac{1}{2}m_{(p)}|\dot{\vec{r}}_{(p)}|\nonumber\\
&-&e\,q_{(p)}\dot{r}_{(p)}^{i}A_{i}(\vec{r}_{(p)},t)-e\,q_{(p)}A_{o}(\vec{r}_{(p)},t)\Big]\nonumber\\
&+&\frac{\kappa}{2}\int
d^{3}x\;\varepsilon^{\mu\nu\lambda}\partial_{\nu}A_{\lambda}(x)A_{\mu}(x)\;.
\label{ec19.a}
\end{eqnarray}
This theory has been studied throughly \cite{4}, mainly due to its
relationship with anyonic statistics. Our main concern will be to
discuss its L.R. formulation. To this end we need the results of the
Dirac quantization of this model, which may be summarized as follows
\cite{4}. The first class Hamiltonian is given by:

\begin{equation}
H=\sum_{p=1}^{N}\frac{1}{2m_{(p)}}\left(\vec{p}_{(p)}-e\,q_{(p)}\vec{A}(\vec{r}_{(p)},t)\right)^{2}\;.
\label{ec20}
\end{equation}
It should be recalled that the Chern-Simons term, due to its topological
character, does not contribute to the energy momentum tensor. That is
why the Hamiltonian in the present case looks like that of a collection
of particles in an external field. Another difference with the previous
case is the commutator

\begin{equation}
\left[A_{i}(\vec{x}), A_{j}(\vec{y})\right]=\frac{i}{\kappa}
\varepsilon^{ij}\delta^{2}(\vec{x}-\vec{y})\;,
\label{ec21}
\end{equation}
which, together with the commutators of the canonical operators for free
particles [i.e., equation (\ref{dificil})] complete the non trivial part
of the algebra of the quantum theory [the remaining commutators vanish
identically]. The first class constraint that replaces the Gauss law of
Maxwell theory, and generates time independent gauge transformations is
given by

\begin{equation}
\kappa
\vec{B}(\vec{x})+\sum_{(p)}e\,q_{(p)}\delta^{2}(\vec{x}-\vec{r}_{(p)})\approx
0\;,
\label{ec23}
\end{equation}
where $B(\vec{x})\equiv -\frac{1}{2} \varepsilon^{ij}F_{ij}$ is the
``magnetic field''. This constraint states that on the physical sector
of the Hilbert space, every particle carries an amount of ``magnetic''
flux proportional to its electric charge, and  confined to the point
where the particle is.

It can be verified that the position and velocity operators,
$\vec{r}_{(p)}$ and $m_{(p)}\vec{v}_{(p)}\equiv
\vec{p}_{(p)}-e\,q_{(p)}\vec{A}(\vec{r}_{(p)},t)$ are gauge invariant.
Moreover, it can be seen that on the physical sector of the Hilbert
space every observable of the theory may be expressed in terms of them
\cite{4}. Then, our next task is to find a suitable realization of these
operators in a geometric representation. As in the theory of the
previous section, we consider the space of path dependent functionals
$\Psi(\gamma_{\vec{r}_{(p)}}, \vec{r}_{(p)})$. The action of the path
and loop derivatives $\delta_{i}(\vec{x})$, $\Delta_{ij}(\vec{x})$, the
Mandelstam derivative $D_{i}(\vec{r}_{(p)})$, and the form factor
$T^{i}(\vec{x}, \gamma)$ is defined as in the former case. Then it is
easy to see that the prescription:

\begin{equation}
A_{i}(\vec{x})\;\rightarrow
\;\frac{i}{e}\delta_{i}(\vec{x})-\frac{e}{2\kappa}\varepsilon_{ij}T^{j}(\vec{x},
\gamma)
\label{ec24}
\end{equation}
realizes the commutator (\ref{ec21}). From this result we can obtain the
velocity operator as

\begin{eqnarray}
m_{(p)}v_{(p)i}=&-&i\left(\frac{\partial}{\partial
r_{(p)}^{i}}+q_{(p)}\delta_{i}(\vec{r}_{(p)})\right)\nonumber\\
&+&\;\frac{e^{2}}{2\kappa}q_{(p)}\varepsilon_{ij}T^{j}(\vec{r}_{(p)},
\gamma)\nonumber\\
=&-&iD_{i}(\vec{r}_{(p)})+\frac{e^{2}}{2\kappa}\,q_{(p)}\,\varepsilon_{ij}T^{j}(\vec{r}_{(p)},
\gamma)\;,
\label{ec25}
\end{eqnarray}
when acting on ``Faraday's lines'' dependent functionals
$\Psi(\gamma_{\vec{r}_{(p)}}, \vec{r}_{(p)})$. After some calculations
one can compute the following commutators in the path representation

\begin{eqnarray}
\left[m_{(p)}v_{(p)}^{i},
m_{(q)}v_{(q)}^{j}\right]&=&i\varepsilon^{ij}\Big(\delta_{pq}e\,q_{(q)}\,B(\vec{r}_{(q)})\nonumber\\
&&+\frac{e^{2}}{\kappa}q_{(p)}\,q_{(q)}\delta^{2}(\vec{r}_{p}-\vec{r}_{q})\Big)\;,\\
\left[r_{(p)}^{i},
m_{(q)}v_{(q)}^{j}\right]&=&i\,\delta^{ij}\delta_{pq}\;,\\
\left[r_{(p)}^{i}, r_{(q)}^{j}\right]&=&0\;,
\end{eqnarray}
and check that they agree with what it is obtained when the same
commutators are calculated directly from the canonical ones, i.e., from
equations (\ref{ec21}) and (\ref{dificil}) \cite{4}.

Our next step will consist on studying the gauge constraint
(\ref{ec23}). Substituting equation (\ref{ec24}) into equation
(\ref{ec23}), we find

\begin{eqnarray}
\left\{\frac{i}{2e}\varepsilon^{ij}\Delta_{ij}(\vec{x})+\frac{e}{2\kappa}\sum_{s}\left(\delta^{2}(\vec{x}-\vec{a}_{s})-\delta^{2}(\vec{x}-\vec{b}_{s})\right)\right.\nonumber\\
\left.\qquad\qquad-\frac{e}{\kappa}\sum_{p=1}^{N}q_{(p)}\delta^{2}(\vec{x}-\vec{r}_{(p)})\right\}\;\Psi(\gamma_{\vec{r}_{(p)}},
\vec{r}_{(p)})=0\, .\nonumber\\
\label{ec29}
\end{eqnarray}
The first two terms of this expression come from the realization
of the magnetic field that rises from equation (\ref{ec24}). There
is a special situation in which one knows the solution of the
path-dependent differential equation (\ref{ec29}), namely, the
case when the charge is proportional to the number of strands

\begin{equation}
q_{(p)}=\alpha\; n_{(p)}\, .
\label{ec30}
\end{equation}
In this case, equation (\ref{ec29}) can be cast in the form

\begin{eqnarray}
&&\left\{\frac{e}{2\kappa}(2\alpha
-1)\sum_{s}\left(\delta^{(2)}(\vec{x}-\vec{b}_{s})-\delta^{(2)}(\vec{x}-\vec{a}_{s})\right)\right.\nonumber\\
&&\left.\qquad\qquad\qquad +\;
\frac{i}{2e}\varepsilon^{ij}\Delta_{ij}(\vec{x})\right\}\,\Psi(\gamma_{\vec{r}_{(p)}},
\vec{r}_{(p)})=0\, ,\nonumber\\
\label{ec30a}
\end{eqnarray}
which we recognize as the first class constraint of the abelian
Maxwell-Chern-Simons theory in an open-path representation
\cite{10}. There is a subtlety which does not spoil the similarity
between the constraints of both theories: in the present study the
points $\vec{r}_{(p)}$ are ``ocupied'' by two entities, the
charged particles that may be displaced by means of
$\partial/\partial r^{i}_{(p)}$, and the boundaries of the paths
that respond to the action of the path derivative
$\delta_{i}(\vec{r}_{(p)})$. In the Maxwell-Chern-Simons case, on
the other hand, there only exist objects of the second type.

The solution of (\ref{ec30a}) is given by \cite{10}

\begin{equation}
\Psi(\gamma_{\vec{r}_{(p)}}, \vec{r}_{(p)})=exp\left(i\frac{e^2
(2\alpha-1)
}{4\pi\kappa}\Delta\Theta(\gamma)\right)\Phi(\partial\gamma_{\vec{r}_{(p)}},
\vec{r}_{(p)}), 
\label{otra30}
\end{equation}
where $\Phi(\partial\gamma_{\vec{r}_{(p)}}, \vec{r}_{(p)})$ is a
function that depends on the path $\gamma_{\vec{r}_{(p)}}$ only through
its boundary $\partial\gamma_{\vec{r}_{(p)}}$, and
$\Delta\Theta(\gamma)$ is the sum of the angles subtended by the pieces
of the path $\gamma$ from their final points $\vec{b}_{s}$, minus the
sum of the angles subtended by these pieces measured from their starting
points $\vec{a}_{s}$:

\begin{eqnarray}
\Delta\Theta(\gamma)\equiv\sum_{s}\int_{\gamma}dx^{k}\,\varepsilon^{lk}\left[\frac{(x-b_{s})^{l}}{|\vec{x}-\vec{b}_{s}|^{2}}-\frac{(x-a_{s})^{l}}{|\vec{x}-\vec{a}_{s}|^{2}}\right]\;.
\label{ec31}
\end{eqnarray}

At this point one should  verify whether the gauge invariant
operators of the theory preserve the form of the physical states
given by equation (\ref{otra30}). It is found that this is so,
provided that $\alpha=1$. For instance, one has for the velocity
operator

\begin{eqnarray}
&&m_{(p)}v_{(p)i}\left[exp\left(i\frac{e^2}{4\pi\kappa}\Delta\Theta(\gamma)\right)\Phi(\partial\gamma_{\vec{r}_{(p)}},
\vec{r}_{(p)})\right]\nonumber\\
&&=exp\left(i\frac{e^2}{4\pi\kappa}\Delta\Theta(\gamma)\right)\times\nonumber\\
&&\quad\Bigg\{\frac{q_{(p)}e^{2}}{2\pi\kappa}\sum_{s}\left[\frac{(r_{(p)}-b_{s})^{i}}{|\vec{r}_{(p)}-\vec{b}_{s}|^{2}}-\frac{(r_{(p)}-a_{s})^{i}}{|\vec{r}_{(p)}-\vec{a}_{s}|^{2}}\right]\nonumber\\
&&\qquad\qquad\qquad\qquad\qquad\quad\quad-\;i\varepsilon^{ij}D_{j}(\vec{r}_{(p)})\Bigg\}\nonumber\\
&&\quad\times\;\Phi(\partial\gamma_{\vec{r}_{(p)}},
\vec{r}_{(p)})\nonumber\\
&&=exp\left(i\frac{e^2}{4\pi\kappa}\Delta\Theta(\gamma)\right)\Phi'(\partial\gamma_{\vec{r}_{(p)}},
\vec{r}_{(p)})\,,
\label{ec32}
\end{eqnarray}
where $\Phi'$, in the last line, is a boundary-dependent
functional, likewise $\Phi$. Hence, we find that a consistent
solution of the gauge constraint is given by equation
(\ref{otra30}), in the case where the charges of the particles
coincide with their number of attached strands. As in the
Maxwell-Chern-Simons case \cite{10} there is an unitary
transformation that allows us to eliminate the path dependent
phase $\chi(\gamma)\equiv
i\frac{e^2}{4\pi\kappa}\Delta\Theta(\gamma)$. It is given by:

\begin{eqnarray}
\Psi(\gamma, \vec{r})&\rightarrow&\widetilde{\Psi}(\partial\gamma,
\vec{r})=exp\left[-\chi(\gamma)\right]\Psi(\gamma,
\vec{r})\,,\\A\,&\rightarrow&\,\widetilde{A}=exp\left[-\chi(\gamma)\right]\,A\;
exp\left[\chi(\gamma)\right]\,,
\label{ec33}
\end{eqnarray}
with $A$ being any gauge invariant operator of the theory. Once
this transformation is performed, the path dependence of the wave
functional $\widetilde{\Psi} $ is reduced to the boundary
$\partial\gamma_{\vec{r}_{(p)}}$ of the path, which is just the
set $\{\vec{r}_{(p)}\}$ of points occupied by the particles.

A moments thought leads one to realize that, at this point, the boundary
dependence of the wave functional becomes redundant, and it suffices to
employ ordinary wave functions $\Psi(\vec{r}_{(p)})$, instead of the
``boundary dependent'' functionals $\Psi(\partial\gamma_{\vec{r}_{(p)}}
, \vec{r}_{(p)})$. At the same time, we should replace the Mandelstam
derivative $D_{i}(\vec{r}_{(p)})=\frac{\partial}{\partial
r_{(p)}^{i}}+q_{(p)}\,\delta_{i}(\vec{r}_{(p)})$ by the ordinary
``point'' derivative $\frac{\partial}{\partial r_{(p)}^{i}}$. The
Schr\"odinger equation of the model may then be written down as

\begin{eqnarray}
i\,\partial_{t}\psi(\vec{r}_{(p)},
t)=\left[\frac{1}{2}\sum_{p=1}^{N}m_{(p)}{v}_{(p)}^{2}\right]\;\psi(\vec{r}_{(p)},
t)
\label{ec34}
\end{eqnarray}
with $m_{(p)}v_{(p)}^{i}$ given by

\begin{equation}
m_{(p)}v_{(p)}^{i}=p_{(p)}^{i}-eq_{(p)}\frac{1}{2\pi\kappa}\varepsilon^{ij}\sum_{q\neq
p}eq_{(q)}\frac{(r_{(p)}^{j}-r_{(q)}^{j})}{|\vec{r}_{(p)}-\vec{r}_{(q)}|^{2}}\;,
\label{ec35}
\end{equation}
and then we recover the well known description of the quantum mechanics
of non-relativistic particles interacting through a quantized
Chern-Simons field \cite{4} that gives rise to a model of anyons. This
fact should be seen as the basic justification for choosing the charge
quantization scheme that we adopted in the Chern-Simons case.

\section{Conclusion}

We have studied the L.R. quantization of point particles
interacting by means of Maxwell and Chern-Simons fields. In both
cases we found that the appropriate Hilbert space is made of wave
functionals whose arguments are Faraday's lines emanating from or
ending at the particles positions. In the Maxwell case, since the
lines of force carry an amount of electric flux that must be a
multiple of the coupling constant $e$, we find that electric
charge must be quantized in order to have a consistent
formulation. In the Chern-Simons case, on the other hand, the
quantization of the electric charge allows to relate, in a simple
form, the geometric representation of the model with the quantum
mechanics of anyons as discussed in references \cite{4,5}. We
think that this feature justifies the choice of the charge
quantization prescription to solve the gauge constraint
(\ref{ec30a}). Hence, in the Chern-Simons case, we obtain the
following picture: the paths may be ``erased'' by means of an
unitary transformation if we prescribe that the charge is
quantized.

We want  to underline how gauge invariance is maintained within
the geometrical framework we have presented. For instance, the
covariant momentum is a generalized derivative, that translates
both the charges and their associated bundles of force lines. In
a similar manner, every gauge invariant operator respects the
geometrical setting where the theory is represented.

It seems possible to develop a similar formulation for models of
extended objects interacting through abelian $p$-forms. It would also be
interesting to explore whether or not charge quantization is necessary
for the consistence of the L.R. of the model of charged fields (instead
of particles) in electromagnetic interaction.

\end{multicols}

\end{document}